\documentclass[12pt,preprint]{aastex61}
\usepackage{amsmath}

\received{2019 March 20}
\revised{2019 April 25}
\accepted{2019 April 29}
\submitjournal{ApJ}

\shorttitle{Detection of 4765$\,$MHz OH Emission in CRL$\,618$}
\shortauthors{Strack, et al.}

\newcommand {\kms }{$\,$km s$^{-1}$}

\begin{document}

\correspondingauthor{E. D. Araya}
\email{ed-araya@wiu.edu}

\title{Detection of 4765$\,$MHz OH Emission in a Pre-Planetary Nebula -- CRL$\,$618}

\author{A. Strack}
\affiliation{Physics Department, Western Illinois University, 
1 University Circle, Macomb, IL 61455, USA.}

\author{E. D. Araya}
\affiliation{Physics Department, Western Illinois University, 
1 University Circle, Macomb, IL 61455, USA.}

\author{M. E. Lebr\'on}
\affiliation{Department of Physical Sciences, University of Puerto Rico, San Juan, Puerto Rico, PR 00931, USA.}

\author{R. F. Minchin}
\affiliation{USRA, SOFIA, NASA Ames Research Center, Moffett Field, CA 94035, USA.}
\affiliation{Arecibo Observatory, NAIC, HC3 Box 53995, Arecibo, Puerto Rico, PR 00612, USA.}

\author{H. G. Arce}
\affiliation{Department of Astronomy, Yale University, New Haven, CT 06511, USA.}

\author{T. Ghosh}
\affiliation{Green Bank Observatory, P.O.Box 2, Green Bank, WV 24944, USA.}
\affiliation{Arecibo Observatory, NAIC, HC3 Box 53995, Arecibo, Puerto Rico, PR 00612, USA.}

\author{P. Hofner}
\altaffiliation{Adjunct Astronomer at the National Radio Astronomy Observatory, 
1003 Lopezville Road, Socorro, NM 87801, USA.}
\affiliation{New Mexico Institute of Mining and Technology, Physics Department, 
801 Leroy Place, Socorro, NM 87801, USA.}

\author{S. Kurtz} 
\affiliation{Instituto de Radioastronom\'{i}a y Astrof\'{i}sica,
Universidad Nacional Aut\'{o}noma de M\'{e}xico, Apdo. Postal 3-72, 
58089, Morelia, Michoac\'{a}n, Mexico.}

\author{L. Olmi} 
\affiliation{INAF, Osservatorio Astrofisico di Arcetri, Largo E. 
Fermi 5, I-50125 Firenze, Italy.}

\author{Y. Pihlstr\"{o}m}
\altaffiliation{Adjunct Astronomer at the National Radio Astronomy Observatory, 1003 Lopezville Road, Socorro, NM 87801, USA.}
\affiliation{Department of Physics and Astronomy, The University of New Mexico, Albuquerque, NM 87131, USA.}

\author{C. J. Salter}
\affiliation{Green Bank Observatory, P.O.Box 2, Green Bank, WV 24944, USA.}
\affiliation{Arecibo Observatory, NAIC, HC3 Box 53995, Arecibo, Puerto Rico, PR 00612, USA.}

\begin{abstract}

Jets and outflows are ubiquitous phenomena in astrophysics, found in our Galaxy in diverse environments, from the formation of stars to late-type stellar objects. We present observations conducted with the 305m Arecibo Telescope of the pre-planetary nebula CRL$\,$618 (Westbrook Nebula) $-$ a well studied late-type star that has developed bipolar jets. The observations resulted in the first detection of 4765$\,$MHz OH in a late-type stellar object. The line was narrow (FWHM $\sim$ 0.6\kms) and $\sim$40\kms~blueshifted with respect to the systemic velocity, which suggests association with the expanding jets/bullets in CRL$\,$618. We also report non-detection at Arecibo of any other OH transition between 1 and 9$\,$GHz. The non-detections were obtained during the observations in 2008, when the 4765$\,$MHz OH line was first discovered, and also in 2015 when the 4765$\,$MHz OH line was not detected. Our data indicate that the 4765$\,$MHz OH line was a variable maser. Modeling of the 4765$\,$MHz OH detection and non-detection of the other transitions is consistent with the physical conditions expected in CRL$\,$618. The 4765$\,$MHz OH maser could originate from dissociation of H$_2$O by shocks after sublimation of icy objects in this dying carbon-rich stellar system, although other alternatives such as OH in an oxygen-rich circumstellar region associated with a binary companion are also possible.

\end{abstract}


\section{INTRODUCTION}

The transition from the asymptotic giant branch (AGB) to the planetary nebula (PN) phase is among the most important areas of study in stellar astrophysics today. The post-AGB/pre-planetary nebula (PPN) phase occurs when a significant fraction of the stellar mass is ejected within a few thousand years, revealing the core of the star, i.e., a new white dwarf (e.g., \citealt{Sanchez_Contreras2017AA...603A..67S}, \citealt{Matthews_2018ASPC..517..281M}). The PPN phase is when asymmetries develop and grow in the structure of the nebula, resulting in a bipolar morphology. The mechanisms responsible for the development of bipolar asymmetries are under investigation and include the effects of rotation, latitudinal density anisotropy, binarity, and magnetic collimation (e.g, \citealt{Garcia-Segura_1999ApJ...517..767G, Garcia-Segura_2014ApJ...783...74G, Blackman_2001Natur.409..485B}). A key probe for the formation and growth of asymmetries has been the study of masers, such as in water fountain objects (e.g., \citealt{Amiri_2010A&A...509A..26A}). Stellar masers of SiO, H$_2$O, and ground state OH are commonly detected in oxygen-rich AGB stars, but during the transition to PNe these masers wither away; first SiO, then H$_2$O and finally the OH masers disappear (e.g., \citealt{Uscanga_2012AA...547A..40U}, \citealt{Desmurs_2012IAUS..287..217D}, \citealt{Habing_1996A&ARv...7...97H}). At present, only seven PNe have been confirmed to show ground state OH masers (\citealt{Qiao_2016ApJ...817...37Q}; \citealt{Uscanga_2012AA...547A..40U}).

A prototypical example of a PPN is CRL 618 (IRAS$\,$04395+3601, V353 Aur, the Westbrook Nebula; \citealt{Cerrigone_2011MNRAS.412.1137C}; \citealt{Trammell_2002ApJ...579..688T}), which is located at a distance of $\sim 900\,$pc (\citealt{Lee_2013ApJ...770..153L}; \citealt{Sanchez_Contreras2004ApJ...617.1142S})\footnote{However, a distance of 1.7$\,$kpc has also been used in the literature (\citealt{Wyrowski_2003ApJ...586..344W}, \citealt{Martin-Pintado_1988A&A...197L..15M}).}. CRL$\,$618 began its post-AGB phase just over 100 years ago (\citealt{Sanchez_Contreras2004ApJ...617.1142S}; \citealt{Kwok_1981ApJ...247L..67K}). The central star (T$_{ef\!fective} \sim 30,000\,$K) has begun ionization of the envelope in the transition from AGB to the PN phase (e.g., \citealt{Pardo_2004ApJ...615..495P}). Variable radio continuum has been detected in this source, which indicates episodes of expansion of the ionization front in the circumstellar envelope (\citealt{Cerrigone_2011MNRAS.412.1137C}; \citealt{Kwok_1981ApJ...247L..67K}). Molecular spectral line observations have shown evidence for at least two significant episodes of mass loss during the past $\sim$2,500 years \citep{Sanchez_Contreras2004ApJ...617.1142S}. CRL$\,$618 is characterized by spherical envelopes, jets/outflows and shocks (bullets), all with different expansion velocities. In particular, it harbors a fast outflow with velocities from 40 to $\sim$300\kms~ (see Figures~5 and 8 in \citealt{Sanchez_Contreras2004ApJ...617.1142S}; see also \citealt{Trammell_2002ApJ...579..688T}).

CRL$\,$618 is a carbon-rich object \citep{Herpin_2001ESASP.460..249H}, i.e., the progenitor AGB was likely a carbon-rich star similar to IRC+10$^\circ$216 and not an OH/IR star. It is rich in molecular emission, including some oxygen species, and for which \citet{Bujarrabal_1988A&A...204..242B} reported the first detection of HCO$^{+}$ in a carbon-rich stellar object, as well as transitions of many molecular species including CO, $^{13}$CO, CS, SiO, CH$_{3}$CN, HC$_3$N, and HC$_5$N. CRL$\,$618 was also the first carbon-rich object with detection of H$_2$CO \citep{Cernicharo1989A&A...222L...1C}.

Regarding the maser environment, \cite{Yoon_2014ApJS..211...15Y} reported no detection of 22$\,$GHz H$_2$O and 43$\,$GHz SiO masers. Until our work, hydroxyl (OH) masers had not been detected either: \cite{Desmurs_2010AA...520A..45D} reported no detection of the 6$\,$GHz OH excited lines, while \cite{Silverglate_1979AJ.....84..345S} reported no detection of the 1612, 1665 or 1667$\,$MHz OH lines. The absence of masers from oxygen-species is not surprising because of the carbon-rich nature of CRL$\,$618. However, using the 305$\,$m Arecibo Telescope, we have detected a narrow line corresponding to the 4765$\,$MHz transition of OH. The detection of this line was a by-product of a survey for H$_2$CO masers reported in \cite{Araya_2015ApJS..221...10A}. In this work we focus on the 4765$\,$MHz OH detection, and present complementary observations also conducted with the Arecibo Telescope.\footnote{An initial report of this work \citep{Strack_2018IAUS..336..385S} was published in the proceedings of the IAU 336 Symposium - Astrophysical Masers: Unlocking the Mysteries of the Universe.}

\section{OBSERVATIONS AND DATA REDUCTION}

Observations of CRL$\,$618 were conducted in 2008, 2015 and 2017 using the Arecibo Telescope. The pointing position was RA = 04$^\text{h}$42$^\text{m}$53\fs64, Decl. = +36\arcdeg06\arcmin53.4\arcsec~(J2000). Channel widths and RMS values of the observations are listed in Tables \ref{table_4765OH_line_par} and \ref{tb_upperlim}. The following subsections give details of the specific runs.

\subsection{2008 May Observations}

The 2008 May observations were conducted as part of a survey for 6$\,$cm H$_2$CO masers toward low-mass star forming regions and late-type stellar objects \citep{Araya_2015ApJS..221...10A}. In addition to the H$_2$CO line, we observed the 4660, 4750 and 4765$\,$MHz transitions of OH and detected 4765$\,$MHz OH emission in CRL$\,$618. Observation and calibration details are given in \cite{Araya_2015ApJS..221...10A}.

\subsection{2008 October Observations}

All OH transitions between 1 and 9 GHz were observed with the Arecibo Telescope during October 2008 using the L-Band Wide, C-Band, C-Band High, and X-Band receivers. The WAPP backend was configured to channel widths between 0.2 and 0.4\kms~for all the observed bands. The observations were made in position-switched mode. The ON-source integration time per scan was 5$\,$min, and three scans were typically obtained per frequency setting. After checking the ON- and OFF-source spectra to search for radio frequency interference (RFI) and signal in the reference position, the spectra were calibrated to flux density units (Jy) using the Arecibo AOIDL package. The orthogonal polarizations were averaged to reduce the noise of the spectra and the line parameters were measured in cases of detection.

\subsection{2015 January, February and October Observations}

The 2015 observations were made to check for variability of the 2008 detection. In addition, all other OH transitions between 1 and 9$\,$GHz were re-observed with the Arecibo Telescope using the L-Band Wide, C-Band, C-Band High, and X-Band receivers. The observations were conducted with the WAPP backend, using channel widths between 0.1 and 0.3\kms~and bandwidths between 1.56$\,$MHz (L-Band) and 6.25$\,$MHz (X-Band). Single 5$\,$min ON-source scans were typically conducted; multiple scans of some spectral windows were obtained to improve sensitivity. Spectra were checked for RFI and calibrated using the Arecibo AOIDL libraries. Polarizations were averaged to reduce the noise of the spectra. 

\subsection{2017 March Observations}

Further monitoring observations of C-Band OH transitions (4660, 4750, and 4765$\,$MHz) were conducted in March 2017. The observations were carried out in position-switched mode (5$\,$min ON-source integration time); five scans were obtained. The WAPP backend was used to observe the three OH lines simultaneously with a bandwidth of 3.12$\,$MHz and 2048 channels in dual linear polarization. The data were calibrated in IDL using the Arecibo AOIDL package. The individual polarizations and scans were inspected for consistency (no polarized signal was detected) and to check for RFI, and were then averaged. The 4765$\,$MHz OH spectrum was smoothed to a channel width of 0.38\kms~(to match the channel width of the October 2008 observations).

\section{RESULTS}

\begin{figure}
\includegraphics{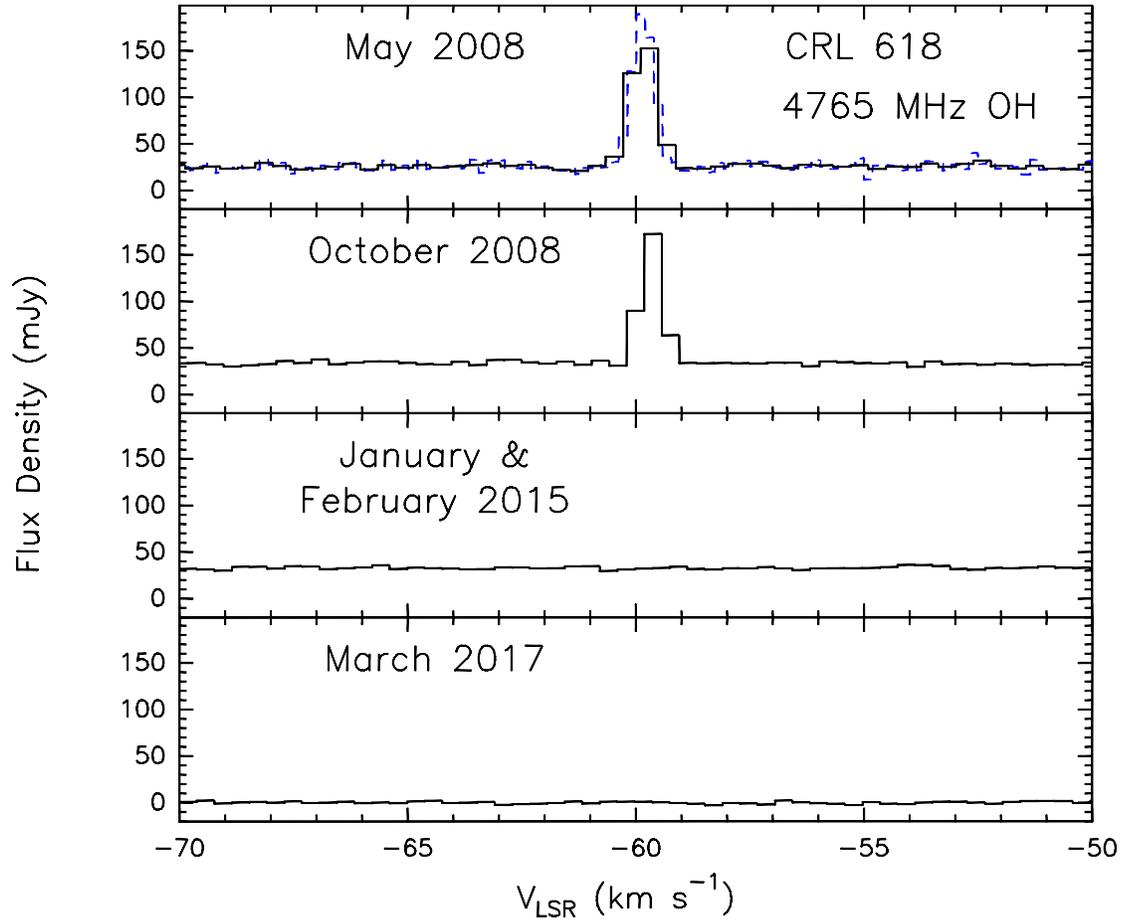}
\vspace*{15.5cm}\caption{The first detection (top panel) and confirmation (middle panel) of 4765$\,$MHz OH emission in a late-type stellar object. The May 2008 observations were conducted with greater spectral resolution than the October 2008 observations. In the top panel, the blue dashed line shows the high spectral resolution data, which were smoothed to generate the spectrum shown in black. The spectral line was not detected in 2015 or 2017. Note the detection of radio continuum from the baseline in 2008 and 2015; as mentioned in Section~3, the 2017 spectrum is shown after baseline subtraction. ASCII files of the spectra shown in this figure (before smoothing) are included as supplemental online materials.}
\label{fig_4765OH}
\end{figure}

The 4765$\,$MHz OH line in CRL$\,$618, obtained as part of an H$_2$CO maser survey by \cite{Araya_2015ApJS..221...10A}, is the first detection of this transition in a late-type stellar object. The 4765$\,$MHz line corresponds to the transition $^2\Pi_{1/2}~ J=1/2~ F=1-0$ ($E_u = 181.94\,$K)\footnote{Spectroscopy information from {\it Splatalogue} (http://www.cv.nrao.edu/php/splat/).}, which is analogous to the 1720$\,$MHz OH ground state line but in the $^2\Pi_{1/2}$ ladder instead of the $^2\Pi_{3/2}$ (e.g., \citealt{Pihlstrom_2008ApJ...676..371P}, see also Figure 5.2 in \citealt{Gray_2012msa..book.....G}). The emission was first detected in the May 2008 observations \citep{Araya_2015ApJS..221...10A}, and then confirmed in October 2008. Figure~\ref{fig_4765OH} shows the 4765$\,$MHz OH spectra from May and October 2008, January/February 2015, and March 2017. Both the May and October 2008 observations showed a detection of 4765$\,$MHz OH emission at V$_{LSR} \sim -$60\kms. The May 2008 observations were conducted with higher spectral resolution than the October 2008 observations; the high resolution (non-smoothed) May 2008 spectrum is shown as a blue dashed line in the upper panel of Figure~\ref{fig_4765OH}. The line parameters of the 4765$\,$MHz OH line are listed in Table~\ref{table_4765OH_line_par}. Table~\ref{tb_upperlim} lists the upper limits of the other OH transitions that were observed, including the RMS noise values, and the channel widths for each of the upper limits. No emission was detected for any of these other OH transitions. 

\begin{deluxetable}{lcccccc}
\tabletypesize{\scriptsize}
\tablecaption{Line Parameters of 4765$\,$MHz OH in CRL$\,$618 \label{table_4765OH_line_par}}
\tablewidth{0pt}
\tablehead{
\colhead{Epoch} & \colhead{Chan. Width} & \colhead{RMS} & \colhead{S$_{\nu, cont}$} & \colhead{S$_{\nu, OH}$} & \colhead{V$_{LSR}$} & \colhead{FHWM} \\
\colhead{} & \colhead{(\kms) } & \colhead{(mJy)} & \colhead{(mJy)} & \colhead{(mJy)} & 
\colhead{(\kms) } & \colhead{(\kms)}}
\startdata
2008 May$^\dag$   & 0.19 & 4.6 & 27      & 166 (4) & $-$59.84 (0.01) & 0.56 (0.02) \\
2008 October$^*$  & 0.38 & 1.5 & 33      & 142 (2) & $-$59.67 (0.01) & 0.58 (0.01) \\
2015 Jan./Feb.    & 0.38 & 1.4 & 28      & \nodata & \nodata         & \nodata     \\  
2015 October      & 0.38 & 1.7 & 29      & \nodata & \nodata         & \nodata     \\  
2017 March        & 0.38 & 1.5 & \nodata & \nodata & \nodata         & \nodata   \\  
\enddata
\tablenotetext{~}{1$\,\sigma$ statistical errors from the fit are 
shown in parentheses. 
$^\dag$ The spectral line is narrow; the line parameters listed here are before smoothing (see Figure~1, upper panel, blue dashed line). If the spectrum is smoothed to the channel width of the 2008 October observations (i.e., $\Delta V = 0.38$\kms), then the line parameters from a Gaussian fit are $S_{\nu, OH} = 145 (4)\,$mJy, $V_{LSR} = -59.85 (0.01)$\kms, $FWHM = 0.63 (0.02)$\kms, $RMS = 2.8\,$mJy.
$^*$ An additional weak spectral line (tentative detection, $\ga 3\sigma$ in two consecutive channels) was also present in the bandpass; the peak channel parameters are $S_{\nu, OH} = 6.6\,$mJy, $V_{LSR} = -39.7$\kms.
}
\end{deluxetable}

\newpage
~\\
\startlongtable
\begin{deluxetable}{lccccc}
\tabletypesize{\scriptsize}
\tablecaption{Upper Limits of other OH Transitions \label{tb_upperlim}}
\tablewidth{0pt}
\tablehead{
\colhead{Rest Freq.} & \colhead{Transition} & \colhead{E$_u$}  & \colhead{Epoch} & \colhead{Chan. Width} & \colhead{RMS} \\
\colhead{(MHz)}      & \colhead{ }          & \colhead{(K)}    & \colhead{ }     & \colhead{(\kms)}      & \colhead{(mJy)}}
\startdata
1612.2310 & $^2\Pi_{3/2}~ J=3/2~ F=1-2$ & 0.080     &  Feb 2015       &  0.14      &  3.9 \\   
           &                            &           &  Oct 2008       &  0.28      &  3.2 \\   
1665.4018 & $^2\Pi_{3/2}~ J=3/2~ F=1-1$ & 0.080     &  Feb 2015       &  0.13      &  3.4 \\   
           &                            &           &  Oct 2008       &  0.28      &  3.1 \\   
1667.3590 & $^2\Pi_{3/2}~ J=3/2~ F=2-2$ & 0.083     &  Feb 2015       &  0.13      &  3.3 \\   
           &                            &           &  Oct 2008       &  0.28      &  3.1 \\   
1720.5300 & $^2\Pi_{3/2}~ J=3/2~ F=2-1$ & 0.083     &  Feb 2015       &  0.13      &  3.3 \\   
           &                            &           &  Oct 2008       &  0.27      &  3.2 \\   
4660.2420 & $^2\Pi_{1/2}~ J=1/2~ F=0-1$ & 181.9     &  May 2008       &  0.20      &  4.4 \\   
           &                            &           &  Oct 2008       &  0.39      &  2.2 \\   
           &                            &           &  Jan/Feb 2015   &  0.19      &  2.7 \\   
           &                            &           &  Oct 2015       &  0.19      &  2.1 \\   
           &                            &           &  Mar 2017       &  0.19      &  1.9 \\
4750.6560 & $^2\Pi_{1/2}~ J=1/2~ F=1-1$ & 181.9     &  May 2008       &  0.19      &  4.6 \\   
           &                            &           &  Oct 2008       &  0.39      &  1.8 \\   
           &                            &           &  Jan/Feb 2015   &  0.19      &  2.7 \\   
           &                            &           &  Oct 2015       &  0.19      &  2.2 \\   
           &                            &           &  Mar 2017       &  0.19      &  1.9 \\
6016.7460 & $^2\Pi_{3/2}~ J=5/2~ F=2-3$ & 120.8     &  Oct 2008       &  0.30      &  2.7 \\   
           &                            &           &  Jan 2015       &  0.23      &  14  \\   
6030.7483 & $^2\Pi_{3/2}~ J=5/2~ F=2-2$ & 120.8     &  Oct 2008       &  0.30      &  2.7 \\   
           &                            &           &  Jan 2015       &  0.23      &  10  \\   
6035.0934 & $^2\Pi_{3/2}~ J=5/2~ F=3-3$ & 120.8     &  Oct 2008       &  0.30      &  2.7 \\   
           &                            &           &  Jan 2015       &  0.23      &  9.5 \\   
6049.0840 & $^2\Pi_{3/2}~ J=5/2~ F=3-2$ & 120.8     &  Oct 2008       &  0.30      &  2.7 \\   
           &                            &           &  Jan 2015       &  0.23      &  10  \\   
7749.9134 & $^2\Pi_{1/2}~ J=3/2~ F=1-2$ & 270.1     &  Oct 2008       &  0.24      &  3.4 \\   
           &                            &           &  Jan 2015       &  0.24      &  18  \\   
7761.7470 & $^2\Pi_{1/2}~ J=3/2~ F=1-1$ & 270.1     &  Oct 2008       &  0.24      &  3.9 \\   
           &                            &           &  Jan 2015       &  0.24      &  14  \\   
7820.1250 & $^2\Pi_{1/2}~ J=3/2~ F=2-2$ & 270.1     &  Oct 2008       &  0.23      &  3.5 \\   
           &                            &           &  Jan 2015       &  0.23      &  8.4 \\   
7831.9552 & $^2\Pi_{1/2}~ J=3/2~ F=2-1$ & 270.1     &  Oct 2008       &  0.23      &  3.5 \\   
           &                            &           &  Jan 2015       &  0.23      &  11  \\   
8118.0567 & $^2\Pi_{1/2}~ J=5/2~ F=2-3$ & 415.9     &  Oct 2008       &  0.23      &  3.8 \\   
           &                            &           &  Jan 2015       &  0.22      &  16  \\   
8135.8649 & $^2\Pi_{1/2}~ J=5/2~ F=2-2$ & 415.9     &  Oct 2008       &  0.22      &  3.8 \\   
           &                            &           &  Jan 2015       &  0.22      &  14  \\   
8189.5856 & $^2\Pi_{1/2}~ J=5/2~ F=3-3$ & 415.9     &  Oct 2008       &  0.22      &  4.5 \\   
           &                            &           &  Jan 2015       &  0.22      &  12  \\   
8207.3939 & $^2\Pi_{1/2}~ J=5/2~ F=3-2$ & 415.9     &  Oct 2008       &  0.22      &  3.9 \\   
           &                            &           &  Jan 2015       &  0.22      &  12  \\   
8503.1503 & $^2\Pi_{1/2}~ J=11/2~ F=5-6$& 1186.7    &  Oct 2008       &  0.22      &  4.4 \\ %  N = 6+ - 6-, J = 11/2 - 11/2, F = 5 - 6   
           &                            &           &  Jan 2015       &  0.21      &  9.6 \\   
8534.8375 & $^2\Pi_{1/2}~ J=11/2~ F=6-6$& 1186.7    &  Oct 2008       &  0.21      &  4.7 \\ % N = 6+ - 6-, J = 11/2 - 11/2, F = 6 - 6  
           &                            &           &  Jan 2015       &  0.21      &  9.3 \\   
8580.1304 & $^2\Pi_{1/2}~ J=11/2~ F=5-5$& 1186.7    &  Oct 2008       &  0.21      &  4.8 \\  % N = 6+ - 6-, J = 11/2 - 11/2, F = 5 - 5 
           &                            &           &  Jan 2015       &  0.21      &  15  \\   
8611.8176 & $^2\Pi_{1/2}~ J=11/2~ F=6-5$& 1186.7    &  Oct 2008       &  0.21      &  4.7 \\  %  N = 6+ - 6-, J = 11/2 - 11/2, F = 6 - 5
           &                            &           &  Jan 2015       &  0.21      &  15  \\   
\enddata
\hspace{-0.5cm}
\tablenotetext{~}{Spectra were smoothed to the channel width listed in the fifth column. Spectroscopy information from {\it Splatalogue} (http://www.cv.nrao.edu/php/splat/) and the NIST Lovas catalog (https://physics.nist.gov/cgi-bin/micro/table5/start.pl); see also \cite{Pihlstrom_2008ApJ...676..371P}.}
\end{deluxetable}

As shown in Figure~\ref{fig_4765OH} and Table~\ref{table_4765OH_line_par}, we detected radio continuum toward CRL$\,$618 from the baseline level of the (ON-OFF)/OFF spectra from 2008 and 2015. The 2017 observations were conducted at sunset, which was likely responsible for changes in the baseline levels between the ON- and OFF-source observations, and radio continuum was not reliably measured at that epoch (the 2017 spectrum in Figure~\ref{fig_4765OH} is shown after baseline subtraction).

\section{DISCUSSION}

\subsection{A 4765$\,$MHz OH Maser in a Carbon-Rich Late-Type Star: Detection and Variability}

OH masers had not been detected in CRL$\,$618 until our May 2008 observations. The detection was confirmed in October 2008 (Figure~\ref{fig_4765OH}). At both epochs the signal was only seen in the ON-source position and in both polarizations, leading us to rule out RFI. We find it unlikely that the emission is coming from an unrelated object, as CRL$\,$618 is the only IRAS source within the beam and main sidelobes. Arecibo has the smallest half-power beam width (HPBW) of all single-dish radio telescopes currently available at 5$\,$GHz (HPBW$_{5\,GHz} \approx 1\arcmin$), and false-positives due to nearby sources should not be a problem. This is in contrast to smaller single-dish telescopes, such as the false-positive detection of ground state OH emission in the carbon star V1187 Ori as demonstrated by observations using the Nan\c{c}ay and Toru\'n telescopes \citep{Szczerba_2002AA...381..491S}. The flux density of the May and October 2008 spectral lines were similar (see Table~\ref{table_4765OH_line_par}). However, the 4765$\,$MHz OH line was not detected in the 2015 and 2017 observations. Given the narrow line-width\footnote{Note, e.g., that the HC$_3$N thermal lines in CRL$\,$618 discussed by \cite{Wyrowski_2003ApJ...586..344W} are broader than 2\kms.} (Table~\ref{table_4765OH_line_par}), lack of detection of other OH transitions (Table~\ref{tb_upperlim}), and variability, we can rule out a thermal origin for the line. The flux density of the OH line reported here is similar to the 1665 and 1667 MHz OH maser lines detected toward the silicate carbon star V778 Cyg of about 100$\,$mJy (\citealt{Little-Marenin_1988ApJ...330..828L}; \citealt{Little-Mareni_1994AA...281..451L}). 

Significant variability of both circumstellar and interstellar masers is a well-known phenomenon (e.g., \citealt{Araya_2010ApJ...717L.133A}, \citealt{te_Lintel_Hekkert_1996A&AS..119..459T}). Some circumstellar maser flares of ground state OH transitions have been observed to last from a few months to several years (e.g., \citealt{Etoka_2017MNRAS.468.1703E}, \citealt{Etoka_1997A&A...321..877E}). The highly variable and rare 4765$\,$MHz OH maser reported in this work is similar to the 1720$\,$MHz OH maser detected toward the young PN K$\,$3-35 \citep{Gomez_2009ApJ...695..930G}\footnote{The nature of K3-35 was called into question by \cite{Engels_1985A&A...148..344E}; although it is widely accepted to be a young PN, e.g., \cite{Miranda_1998MNRAS.298..243M}.}, which is one of only three known cases of 1720$\,$MHz OH masers in PNe (\citealt{Gomez_2016MNRAS.461.3259G}; \citealt{Qiao_2016ApJ...817...37Q}). The flux density of the 1720$\,$MHz OH maser in K$\,$3-35 decreased from 3$\,$Jy in 1988 (\citealt{te_Lintel_Hekkert_1991A&A...248..209T}, see their Figure~1o) to 0.6$\,$Jy by 2002 \citep{Gomez_2009ApJ...695..930G}. As in the case of the 4765$\,$MHz OH maser in CRL$\,$618, the single-dish spectrum of the 1720$\,$MHz OH maser in K$\,$3-35 is dominated by a single narrow line.

In the case of carbon-rich objects, H$_2$O masers in silicate carbon stars are also highly variable. For example, the H$_2$O maser in NC83 has shown variability by a factor of more than 40, from non-detections in the 1980s and 1990s, to a 4.1$\,$Jy line in 2007 \citep{Ohnaka-2013AA...559A.120O}. Another example is the silicate carbon star EU And, where \cite{Little-Marenin_1988ApJ...330..828L} and \cite{Benson_1987ApJ...316L..37B} reported the detection of a 22$\,$GHz H$_2$O maser redshifted by 24\kms~with respect to the systemic velocity (see also \citealt{Engels_1994AA...285..497E}). This maser was variable (undetectable at some epochs, \citealt{Little-Marenin_1988ApJ...330..828L}, \citealt{Nakada_1987ApJ...323L..77N}) and its spectrum was dominated by a single narrow (FWHM = 0.6\kms) line, i.e., similar to the OH maser reported in the present work. Variability of HCN masers in carbon stars (including masers that disappeared within tens of days to several months after detection) has also been detected in other sources (e.g., \citealt{Lucas_1988A&A...194..230L}, \citealt{Izumiura_1995ApJ...440..728I}, \citealt{Schilke_Menten_2003ApJ...583..446S}).

The LSR velocity of the 4765$\,$MHz OH line reported here is approximately $-$60\kms, and the systemic velocity of the star is $-21.5$\kms~\citep{Sanchez_Contreras2004ApJ...617.1142S}. Thus, the line is blueshifted by $\sim 40$\kms. Given our detection and the radial velocity profiles discussed by \cite{Sanchez_Contreras2004ApJ...617.1142S}, the maser could be associated with the interface between the fast bipolar outflow and the dense core region (see their Figure~5; also Figure 11 in \citealt{Lee_2013ApJ...770..153L}). \cite{Cernicharo1989A&A...222L...1C} reported a terminal velocity outflow of $\approx 200$\kms~based on CO observations, and detected an HC$_3$N absorption feature at $-57.5$\kms~(i.e., very similar to the 4765$\,$MHz OH velocity). They interpreted this high-velocity feature as tracing absorption against the central radio continuum region.

High velocity lines have been detected in other post-AGB objects with bipolar asymmetries. For instance, early work by \citet{teLintelHekkert_1988A&A...202L..19T} reported examples of highly evolved AGB stars with main line OH transitions detected over a broad velocity range ($\ga 100$\kms), including the water fountain IRAS$\,$16342$-$3814 \citep{Likkel_Morris_1988ApJ...329..914L}, which has shown flux density variability over time periods as short as months with minimal shifts in velocity ($\la 1$\kms; \citealt{{Likkel_1992A&A...256..581L}}). Likewise, in the case of CRL$\,$618, the LSR peak velocity of the two detections agree within 1\kms~(see Table~\ref{table_4765OH_line_par}). In the case of the proto-typical carbon star IRC+10$^\circ$216, a narrow ($FWHM \sim 0.3$\kms) SiS maser line was also detected, blueshifted with respect to the systemic velocity, albeit by a smaller amount ($13.5$\kms, \cite{Nguyen-Q-Rieu_1984ApJ...286..276N}, assuming a systemic velocity of $-26.5$\kms). \citet{Nguyen-Q-Rieu_1984ApJ...286..276N} reported a $\sim 40$\% variability of the SiS ($J = 1-0$) maser over a period of 10 months. Preferentially blueshifted emission with respect to systemic velocity has also been reported in ground state OH masers associated with PNe \citep{Uscanga_2012AA...547A..40U}.

An analogous system to CRL$\,$618 is the bipolar PPN object IRAS$\,$08005$-$2356 \citep{Sahai_2015ApJ...810L...8S}, where 1612$\,$MHz OH masers have been detected \citep{te_Lintel_Hekkert_1991_AAS...90..327T} and there is evidence for dual carbon and oxygen chemistry \citep{Bakker_1997AA...323..469B}. The 1612$\,$MHz OH masers in IRAS$\,$08005$-$2356 are blueshifted from the systemic velocity by $\sim 50$\kms~(\citealt{te_Lintel_Hekkert_1991_AAS...90..327T}; \citealt{Slijkhuis_1991AA...248..547S}). IRAS$\,$07027$-$7934 is yet another example of a carbon-rich young PN with a 1612$\,$MHz OH maser, which may have transformed from an OH/IR star into a carbon star within the past few hundred years \citep{Zijlstra_1991AA...243L...9Z}.

Excited OH lines in late-type stellar objects are extremely rare \citep{Habing_1996A&ARv...7...97H}. The only confirmed masers are toward the PPNe Vy 2-2 and K 3-35 at 6035$\,$MHz (variable masers, \citealt{Desmurs_2010AA...520A..45D}). Two other unconfirmed detections were possible artifacts (AU Gem, 4750$\,$MHz OH) or highly variable masers (NML Cyg, 6030 and 6035$\,$MHz; \citealt{Sjouwerman_2007ApJ...666L.101S} and references therein). In the case of Vy 2-2, the OH line is also significantly blueshifted ($\sim 20$\kms~with respect to systemic; \citealt{Desmurs_2002A&A...394..975D}). We note that 6$\,$cm excited OH lines were not detected in our Arecibo spectral-line survey of IRC+10$^\circ$216 \citep{Araya_2003ApJ...596..556A}. 

In summary, although excited OH lines in late-type stellar objects are rare, the overall characteristics of our detection (narrow line, low flux density, variability, blue-shifted line with respect to the systemic velocity) are similar to other masers in late-type stellar objects.

\subsection{Radio Continuum}

\citet{Tafoya2013A&A...556A..35T} reported long-term monitoring observations of the radio continuum in CRL$\,$618 at 5 and 22$\,$GHz. They found an increase in the 5$\,$GHz radio continuum flux density from $6 \pm 3 \,$mJy in 1974 to $33 \pm 2 \,$mJy in 1998, which they characterized by a +0.8 power law ($S_\nu \propto (t-t_0)^{0.8}$). The 22$\,$GHz continuum variability was also well fit by a +0.8 power law (data between 1982 and 2007). They also found that the angular size of the ionized region monotonically increased and estimated that the circumstellar ionization began in the early 1970's by extrapolating to zero angular size. As shown in Table~\ref{table_4765OH_line_par}, we detected approximately the same 5$\,$GHz flux density as that measured in the 1990's. This suggests that the 5$\,$GHz radio continuum flux density from the ionized gas in this young PPN remained approximately constant between 1998 and 2015, although the +0.8 power law flux density increase is still consistent with our measurement within the uncertainty from the fit reported in Equation~1 of \cite{Tafoya2013A&A...556A..35T}. Further Arecibo and VLA observations are needed to investigate whether the 5$\,$GHz radio continuum has changed since then. We note that according to \cite{Wyrowski_2003ApJ...586..344W}, the VLA radio continuum observations of CRL$\,$618 are not significantly affected by resolved out extended structure.

\begin{figure}
\includegraphics{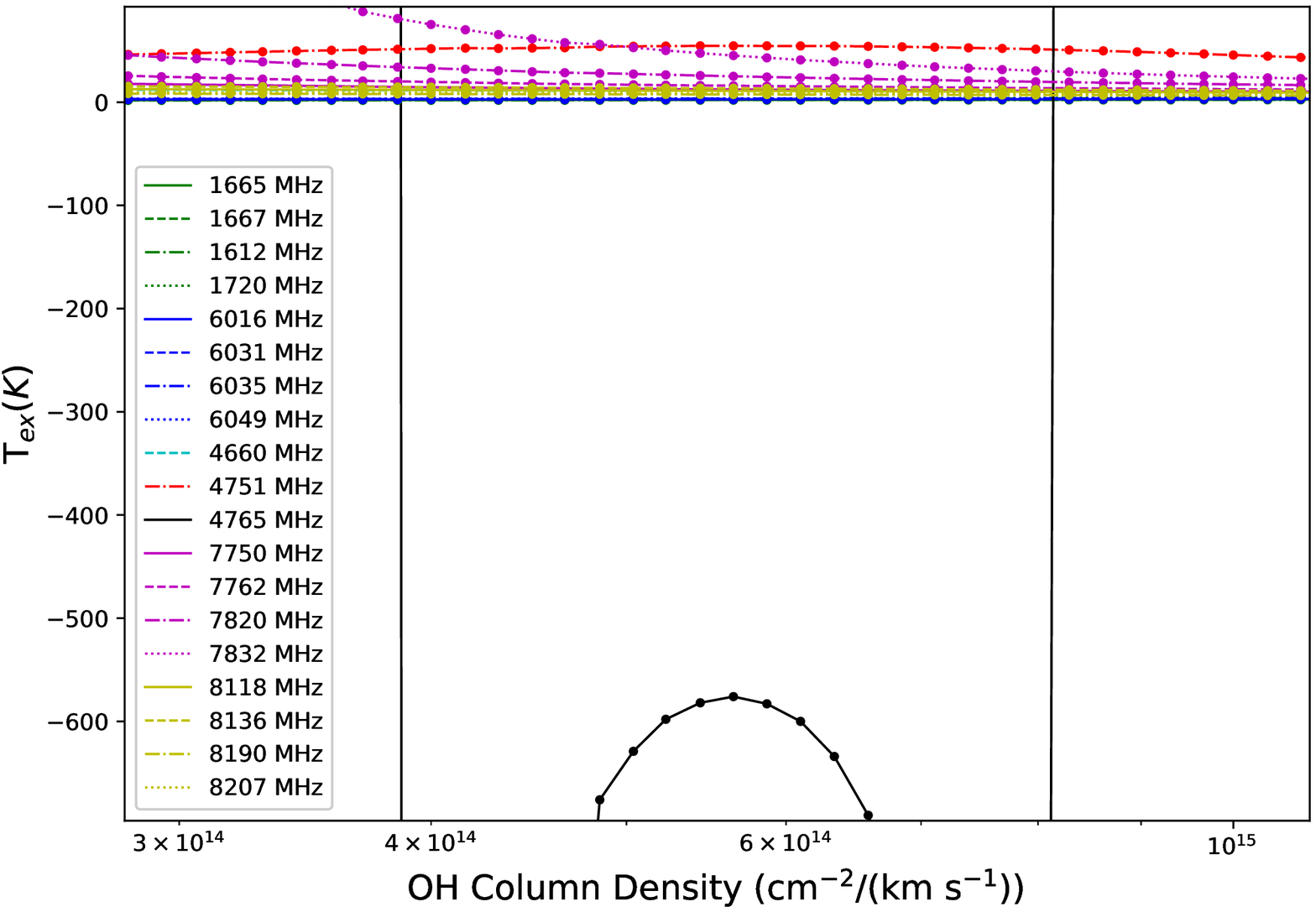}
\includegraphics{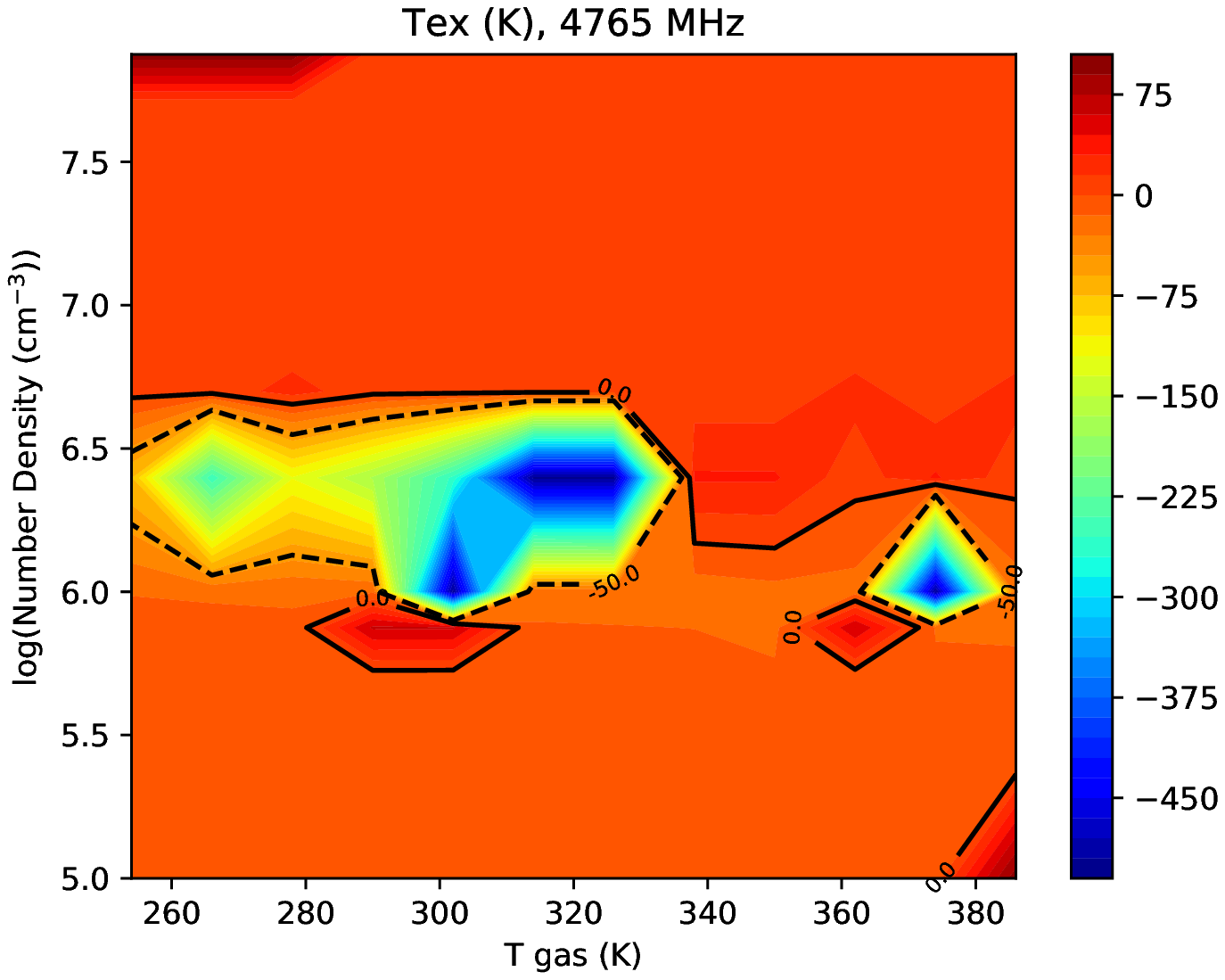}
\vspace*{17.5cm}\caption{Radiative transfer model of OH transitions observed in this work using the code {\tt Molpop} \citep{Elitzur_Asensio_2006MNRAS.365..779E}. {\it Upper panel} shows the excitation temperature of the 4765$\,$MHz OH transition as a function of molecular density and gas temperature. {\it Lower panel} shows that inversion of the 4765$\,$MHz OH line, but not the other transitions, is possible under specific physical conditions (see $\S$\ref{ssec:num1}). The vertical lines mark the range of specific column densities where the 4765$\,$MHz OH transition is inverted for a given set of physical parameters ($\S$\ref{ssec:num1}). The asymptotically negative excitation temperature values approaching the vertical lines mark the transition to non-inverted states where the maser action disappears (e.g., see Figure~7 of \citealt{Thaddeus_1972ApJ...173..317T}).  }
\label{molpop_fig}
\end{figure}

\subsection{Maser Excitation}\label{ssec:num1}

We used the 1-D radiative transfer code {\tt Molpop} \citep{Elitzur_Asensio_2006MNRAS.365..779E} to check whether it is theoretically possible to have realistic physical conditions for which the 4765$\,$MHz OH transition is inverted while the other transitions observed in this work are not (see Table~\ref{tb_upperlim}). {\tt Molpop} has the option of invoking the escape probability approximation for the solution of multi-level radiative transfer problems using a plane-parallel slab approximation. We explored a broad parameter space of densities between $10^1$ and $10^{10}\,$cm$^{-3}$, gas temperatures between 50 and 450$\,$K, radiation field from dust at temperatures between 50 and 200$\,$K, and line overlapping excitation with a linewidth of 1\kms~or 12\kms. As an example, Figure~\ref{molpop_fig} shows the results of models assuming a radiation field from warm dust at 150$\,$K, dust $\tau_V = 100$ \citep{Wyrowski_2003ApJ...586..344W}, and including line overlapping excitation with a linewidth of 12\kms. The {\it upper panel} shows the excitation temperature of the 4765$\,$MHz OH transition within a sub-range of densities and gas temperatures and the {\it lower panel} shows the excitation temperature as a function of OH column density for a gas temperature of 315$\,$K and density of 2.5$\times 10^6\,$cm$^{-3}$. This model predicts population inversion (negative excitation temperature) of only the 4765$\,$MHz OH transition at an OH column density per unit linewidth of $\sim 6\times 10^{14}\,$cm$^{-2}/($\kms$)$. Other values for the model parameters (e.g., a gas temperature of 158$\,$K, number density of $5 \times 10^6\,$cm$^{-3}$ and OH column density per unit linewidth of $\sim$8$\times 10^{14}\,$cm$^{-2}/($\kms$)$) also result in inversion of the 4765$\,$MHz OH energy states, while other transitions show weak or no inversion. Given that the maser was not imaged (its angular size and location with respect to the continuum source are unknown), no attempt was made to model the flux density or brightness temperature of the maser based on the inversion level; our intent is to simply explore whether the 4765$\,$MHz OH transition can be inverted while the other transitions are not. We note that similar models developed for star-forming environments (\citealt{Ellingsen_2004MNRAS.354..401E}; \citealt{Cragg_2002MNRAS.331..521C}; \citealt{Pavlakis_1996ApJ...467..309P}) also predict that specific physical conditions can lead the 4765$\,$MHz OH transition to be a dominant maser line with large brightness temperatures ($> 10^4\,$K).

The conditions used for the model shown in Figure~\ref{molpop_fig} are reasonable based on the complex environment of CRL$\,$618. The rotation temperatures derived from HC$_3$N and HC$_5$N transitions range from 80 to 670$\,$K \citep{Bujarrabal_1988A&A...204..242B}, which encompass the gas temperature of 315$\,$K used in the model. A temperature of $\sim 300\,$K was also assumed in the model of \cite{Pardo_2004ApJ...615..495P} for the slowly expanding molecular gas region at the base of the high velocity wind (see their Figure~4). Also, the chemical structure model presented by \cite{Herpin_Cernicharo_2000ApJ...530L.129H} (e.g., see their Figure~3) locates the OH-rich region inside the high-velocity wind region at a temperature of $\sim 250\,$K. \cite{Silverglate_1979AJ.....84..345S} reported an upper limit of 8.6$\times$10$^{16}$ cm$^{-2}$ for the OH column density in CRL$\,$618, which is consistent with the column density needed for inversion shown in Figure~\ref{molpop_fig}. \cite{Herpin_Cernicharo_2000ApJ...530L.129H} report an OH column density of 8$\times$10$^{15}$ cm$^{-2}$, which is also similar to the one from Figure~\ref{molpop_fig} (note that the abscissa is in units of column density per unit linewidth). The linewidth of 12\kms~assumed for the line-overlap effect is also possible; e.g., \cite{Lee_2013ApJ...770..153L} reported multiple molecular transitions with linewidths $\ge 10$\kms~(see their Figure~9). \cite{Martin-Pintado_1988A&A...197L..15M} modeled the IR SED of CRL$\,$618 as the superposition of two black bodies with temperatures of 95 and 280$\,$K, which encompass the dust temperature of 150$\,$K used to generate Figure~\ref{molpop_fig}. We note that the slowly expanding ($\sim 20$\kms) and relatively cold (30 - 100$\,$K) gas+dust envelope discussed by \cite{Cernicharo1989A&A...222L...1C} could also contribute to the IR field responsible for the excitation. The density assumed in Figure~\ref{molpop_fig} (2.5$\times 10^6\,$cm$^{-3}$) is similar to the density of the high-velocity wind region ($10^5 - 10^6\,$cm$^{-3}$) and the density of the $\sim 20$\kms~expanding molecular torus ($\sim 10^7\,$cm$^{-3}$) reported by \cite{Herpin_Cernicharo_2000ApJ...530L.129H}. Thus, from a theoretical perspective, physical conditions in environments similar to those expected in CRL$\,$618 can lead the 4765$\,$MHz OH transition to be inverted.  

As mentioned above, the 4765$\,$MHz OH detection is blueshifted by $\sim$40\kms~with respect to the systemic velocity, and there is evidence that the OH gas traces the inner circumstellar envelope \citep{Herpin_Cernicharo_2000ApJ...530L.129H}. Thus, the maser is likely located in the foreground of the radio continuum source. This is similar to the case of the very young PN K3-35 where the 1667 and 1612$\,$MHz OH masers are blue-shifted with respect to the systemic velocity by $\sim 20$\kms~\citep{Gomez_2009ApJ...695..930G}. Given the variability of the 4765$\,$MHz OH maser in CRL$\,$618, it is likely that the maser was unsaturated, i.e., exponential growth of flux density with distance through the medium by amplifying the background radio continuum (e.g., \citealt{Gray_2012msa..book.....G}). However, a reliable estimate of the maser gain is not possible because of the uncertain location of the maser with respect to the background radio continuum. The characteristics of the OH maser in CRL$\,$618 are similar to other masers in carbon-rich objects. For example, the HCN (1-0) maser in Y CVn was modeled by \citet{Dinh-V-Trung_2000AA...361..601D} as unsaturated emission (consistent with a high level of variability) and $T_{ex} < -100\,$K.

\subsection{Scenarios for the Origin of the 4765$\,$MHz OH Maser in CRL 618}

In OH/IR stars it is thought that OH is produced through photo-dissociation of H$_{2}$O (e.g., \citealt{Habing_1996A&ARv...7...97H}). In some cases, 1665 and 1667$\,$MHz OH masers have been found located in regions where H$_{2}$O masers originate. For instance, \cite{{Richards_2011A&A...525A..56R}} studied U Ori and U Her that have shown OH and H$_2$O maser variability and where the OH masers may originate from the inner circumstellar envelope \citep{Etoka_1997A&A...321..877E}. In the case of CRL$\,$618, \cite{Herpin_Cernicharo_2000ApJ...530L.129H} detected H$_2$O and OH transitions at far-IR wavelengths.
\cite{Herpin_Cernicharo_2000ApJ...530L.129H} discussed the possibility that OH could be produced through the photo-dissociation of CO and recombination of molecules. The OH and H$_{2}$O produced via this process occupy the innermost molecular envelope where the UV flux is the strongest. Other possible mechanisms that may contribute to the production of OH are shock-induced chemistry and photo-dissociation of H$_{2}$O in the interface between the ionized region and the circumstellar gas, which is pushed at high speeds in the outflow (note that \cite{Tafoya2013A&A...556A..35T} reported expansion of the ionized region in CRL$\,618$). \cite{Cernicharo1989A&A...222L...1C} pointed out that OH and other molecules can reform after $\sim 1\,$year in post-shocked gas (see also \citealt{Neufeld_Dalgarno_1989ApJ...340..869N}), thus, significant velocity shifts with respect to systemic would be expected, as in the case of our detection.

The likely association of the OH maser gas with H$_2$O is also strengthened by the case of the young PN K3-35, where \cite{Gomez_2009ApJ...695..930G} reported detection of masers from the four ground state OH lines. They found 1720$\,$MHz masers to have very similar velocity and location to the H$_2$O masers. An excited OH maser of the 6035$\,$MHz transition has also been detected toward K3-35, and, as for the 1720$\,$MHz OH masers, the excited OH maser in K3-35 is coincident in position and velocity with H$_2$O masers and may originate from the inner circumstellar envelope (\citealt{Gomez_2009ApJ...695..930G}, \citealt{Desmurs_2010AA...520A..45D}).

According to the model by \cite{Herpin_Cernicharo_2000ApJ...530L.129H}, the region where OH and H$_2$O are located in CRL$\,$618 is within 1,400$\,$a.u. of the star. In our Solar System, that volume includes the Kuiper Belt and the inner Oort Cloud where countless icy objects are found. We point out that in the carbon-rich environment of CRL$\,$618, a source of H$_2$O could be sublimation from icy objects in its Kuiper Belt/Oort Cloud; H$_2$O could then be dissociated to form the OH molecules required for the maser. As pointed out by \cite{Stern_1990Natur.345..305S} and \cite{Melnick_2001Natur.412..160M}, H$_2$O from icy objects around an AGB star may be directly sublimated by the thermal radiation of the star up to several hundred astronomical units away from the central source, and thus, icy objects much farther away (in an analogue Oort Cloud) would not be a source of H$_2$O vapor if only thermal sublimation is considered. Instead, taking into account the significant velocity difference between the maser emission and the systemic velocity, shocks interacting with the icy-objects in the outflow could be responsible for the sublimation. As demonstrated by the {\it ESA Rosetta} mission to comet 67P/Churyumov-Gerasimenko, outgassing in a comet can result in H$_2$O column densities between 10$^{14}$ and 10$^{16}\,$cm$^{-2}$ \citep{Marschall_2016A&A...589A..90M}, i.e., similar to the OH column density required by our model. Thus, the detection of the short-lived 4765$\,$MHz OH maser reported in this work may not only be tracing the end of the star's life in its transition to a planetary nebula, but also the destruction of icy objects that have existed since the formation of the star.

CRL$\,$618 is a more-evolved example of systems similar to the carbon star IRC+10$^\circ$216, as CRL$\,$618 has a greater effective temperature ($\sim 32,000\,$K, \citealt{Tafoya2013A&A...556A..35T}, and references therein) than IRC+10$^\circ$216 ($\sim 2,200\,$K, e.g., \citealt{Matthews_2007AJ....133.2291M}), and CRL$\,$618 shows strong radio continuum emission from ionized gas as the star transitions to a PN (e.g., \citealt{Tafoya2013A&A...556A..35T}; the radio continuum in IRC+10$^\circ$216 is significantly weaker, see \citealt{Dinh-V-Trung_2008ApJ...678..303D}). \cite{Melnick_2001Natur.412..160M} reported detection of a 556.9$\,$GHz H$_2$O transition in IRC+10$^\circ$216 and interpreted this detection as evidence for sublimation of H$_2$O from orbiting comets. Ford et al. (2003) later detected 1665 and 1667$\,$MHz OH lines using the Arecibo Telescope (the OH lines in IRC+10$^\circ$216 are blueshifted by $\sim$10\kms~with respect to the systemic velocity). As pointed out by \cite{Ford2003ApJ...589..430F} (and references therein), OH in IRC+10$^\circ$216 could also originate from sublimation of icy objects in a Kuiper belt analog (an extrasolar cometary system).

The idea of destruction of icy objects has also been proposed as the source of OH in silicate carbon stars with OH ground state maser detections (e.g., \citealt{Szczerba_2007BaltA..16..134S}). Other scenarios proposed to explain ground state OH masers in silicate carbon stars include the presence of an oxygen-rich envelope ejected before the carbon enrichment during the thermal pulses and dredge-up, or an oxygen-rich disk around a companion star \citep{Szczerba_2007BaltA..16..134S}. Indeed, water masers in silicate carbon stars are thought to originate in an oxygen-rich circum-binary or circumstellar disk around a low-luminosity companion of the carbon star (a white dwarf or main sequence star; e.g., see \citealt{Ohnaka-2013AA...559A.120O}, \citealt{Little-Marenin_1988ApJ...330..828L}, \citealt{Benson_1987ApJ...316L..37B}, \citealt{Szczerba_2006AA...452..561S}; although a white dwarf companion may be less likely, see \citealt{Deguchi_1988ApJ...325..795D}). \cite{Cernicharo1989A&A...222L...1C} mentioned that the bipolar outflow in CRL$\,$618 could be caused by a close binary, and \cite{Tafoya2013A&A...556A..35T} also discussed the possibility of a binary system based in part on the morphology of the radio continuum. Thus, the OH maser in CRL$\,$618 could have originated instead from an oxygen-rich region similar to those proposed for silicate carbon stars (e.g., \citealt{Ohnaka-2013AA...559A.120O}).

\section{CONCLUDING REMARKS}

We report high sensitivity observations conducted with the 305-m Arecibo Telescope spread over a decade of the 4765$\,$MHz OH line in CRL$\,$618. The line was detected in two independent observing runs in 2008. However, it was not detected in either 2015 or 2017. This is the first detection of the 4765$\,$MHz OH transition in a late-type stellar object, and is a particularly interesting result because CRL$\,$618 is a carbon-rich PPN. Using Arecibo, we observed all other OH transitions between 1 and 9$\,$GHz, but detected no other OH line. Observations of the other transitions were conducted when the 4765$\,$MHz OH line was detected (in 2008) and when that line was not detected (in 2015). The narrow linewidth of the 4765$\,$MHz OH line (0.6\kms), the absence of other OH transitions, the large velocity difference ($\sim$40\kms~blueshifted) between the line and the systemic velocity of the star and its large variability indicate that the 4765$\,$MHz OH line was a maser.

Using the radiative transfer code {\tt Molpop} \citep{Elitzur_Asensio_2006MNRAS.365..779E}, we find that inversion of the 4765$\,$MHz OH line (and absence of all other OH transitions between 1 and 9$\,$GHz) can be explained assuming reasonable physical conditions for a PPN environment. We propose that the OH may originate from sublimation of H$_2$O molecules from icy-objects in the Kuiper Belt/Oort Cloud analogous region of CRL$\,$618, followed by molecular dissociation by UV radiation or shocks. Further monitoring of the 4765$\,$MHz OH transition in CRL$\,$618 and other PPNe is required to investigate how common this line is in PPN objects. If this transition traces the inner regions of PPNe, VLBI observations could be used to investigate the development of bipolar structures.

\acknowledgments

We thank an anonymous referee for comments that improved the manuscript. A.S. and E.D.A. acknowledge the support of WIU Distinguished Alumnus Frank Rodeffer to the WIU Physics Department and the WIU Astrophysics Research Laboratory, in particular student scholarships and computational resources. E.D.A. acknowledges partial support from NSF grant AST-1814063. P.H. acknowledges partial support from NSF grant AST-1814011. Until March 31, 2018, the Arecibo Observatory was operated by SRI International under a cooperative agreement with the National Science Foundation (AST-1100968), and in alliance with Ana G. M\'endez-Universidad Metropolitana, and the Universities Space Research Association. This research has made use of NASA's Astrophysics Data System; the SIMBAD database, operated at CDS, Strasbourg, France; and the NASA/ IPAC Infrared Science Archive, which is operated by the Jet Propulsion Laboratory, California Institute of Technology, under contract with the National Aeronautics and Space Administration.

\software{AOIDL, Molpop \citep{Elitzur_Asensio_2006MNRAS.365..779E}}

\end{document}